\documentclass[sigconf]{_acm/acmart}

\usepackage{bigstrut}
\usepackage{booktabs}
\usepackage{multirow}
\usepackage{microtype}
\usepackage{graphicx}
\usepackage{subfig}
\usepackage{bbding}
\usepackage{overpic}
\usepackage{balance}
\usepackage{amsthm}
\usepackage{array}
\usepackage{clrscode}
\usepackage{multicol}
\usepackage{float}
\usepackage{newfloat}
\usepackage{color}
\usepackage{xcolor}
\usepackage{amsopn}
\usepackage{mathrsfs}
\usepackage{mathtools}
\usepackage{amsmath}
\usepackage{arydshln}
\usepackage{blkarray}
\usepackage{enumerate}
\usepackage{courier}
\usepackage{rotating}
\usepackage{bm}
\usepackage{wrapfig}
\usepackage{ragged2e}
\usepackage[misc]{ifsym}
\usepackage{threeparttable}
\usepackage{makecell}
\usepackage{adjustbox} % For adjusting table size
\usepackage{enumitem}
\usepackage{url}
\usepackage{xspace}
\usepackage{comment}
\usepackage{changepage}
\usepackage{algorithm}
\usepackage{algpseudocode} 
\usepackage{algorithmicx}

\usepackage{amssymb}
\usepackage{utfsym}
\usepackage[ruled, vlined, nofillcomment, linesnumbered, algo2e]{algorithm2e}
\usepackage[normalem]{ulem}
\usepackage{hyperref}
\useunder{\uline}{\ul}{}
\newcommand{\etc}{\emph{etc.}}
\newcommand{\ie}{\emph{i.e.,}\xspace}
\newcommand{\eg}{\emph{e.g.,}\xspace}

\newcommand{\paratitle}[1]{\vspace{1.5ex}\noindent\textbf{#1}}

\newcommand{\ignore}[1]{}

\AtBeginDocument{%
  \providecommand\BibTeX{{%
    \normalfont B\kern-0.5em{\scshape i\kern-0.25em b}\kern-0.8em\TeX}}}

\copyrightyear{2025} 
\acmYear{2025} 
\setcopyright{acmlicensed}\acmConference[KDD '25]{Proceedings of the 31st ACM SIGKDD Conference on Knowledge Discovery and Data Mining V.2}{August 3--7, 2025}{Toronto, ON, Canada}
\acmBooktitle{Proceedings of the 31st ACM SIGKDD Conference on Knowledge Discovery and Data Mining V.2 (KDD '25), August 3--7, 2025, Toronto, ON, Canada}
\acmDOI{10.1145/3711896.3736869}
\acmISBN{979-8-4007-1454-2/2025/08}

\begin{document}

\title{Bridging Textual-Collaborative Gap through Semantic Codes for Sequential Recommendation}

%%
% \settopmatter{authorsperrow=4}
\author{Enze Liu}
\orcid{0009-0007-8344-4780}
\affiliation{%
    \department{Gaoling School of Artificial Intelligence}
    \institution{Renmin University of China}
    \city{Beijing}
    \country{China}
}
\email{enzeliu@ruc.edu.cn}

\author{Bowen Zheng}
\orcid{0009-0002-3010-7899}
\affiliation{%
    \department{Gaoling School of Artificial Intelligence}
    \institution{Renmin University of China}
    \city{Beijing}
    \country{China}
}
\email{bwzheng0324@ruc.edu.cn}

\author{Wayne Xin Zhao\textsuperscript{\Letter}}
\orcid{0000-0002-8333-6196}
\affiliation{
    \department{Gaoling School of Artificial Intelligence}
    \institution{Renmin University of China}
    \city{Beijing}
    \country{China}
}
\email{batmanfly@gmail.com}

\author{Ji-Rong Wen}
\orcid{0000-0002-9777-9676}
\affiliation{
    \department{Gaoling School of Artificial Intelligence}
    \institution{
    Renmin University of China}
    \city{Beijing}
    \country{China}
}
\email{jrwen@ruc.edu.cn}

\thanks{\Letter \ Corresponding author.}

\renewcommand{\shortauthors}{Enze Liu, et al.}

%%
% Abstract.
\begin{abstract}
In recent years, substantial research efforts have been devoted to enhancing sequential recommender systems by integrating abundant side information with ID-based collaborative information. This study specifically focuses on leveraging the textual metadata (e.g., titles and brands) associated with items. While existing methods have achieved notable success by combining text and ID representations, they often struggle to strike a balance between textual information embedded in text representations and collaborative information from sequential patterns of user behavior.
In light of this, we propose \textbf{CCFRec}, a novel \underline{C}ode-based textual and \underline{C}ollaborative semantic \underline{F}usion method for sequential \underline{Rec}ommendation.
The key idea behind our approach is to bridge the gap between textual and collaborative information using semantic codes.
Specifically, we generate fine-grained semantic codes from multi-view text embeddings through vector quantization techniques. Subsequently, we develop a code-guided semantic-fusion module based on the cross-attention mechanism to flexibly extract and integrate relevant information from text representations.
In order to further enhance the fusion of textual and collaborative semantics, we introduce an optimization strategy that employs code masking with two specific objectives: masked code modeling and masked sequence alignment. The merit of these objectives lies in leveraging mask prediction tasks and augmented item representations to capture code correlations within individual items and enhance the sequence modeling of the recommendation backbone. Extensive experiments conducted on four public datasets demonstrate the superiority of CCFRec, showing significant improvements over various sequential recommendation models.
Our code is available at \href{https://github.com/RUCAIBox/CCFRec}{{\textcolor{blue}{https://github.com/RUCAIBox/CCFRec}}}.

\end{abstract}

%%
% ACM Computing Classification System.


%%
% Keywords.


%%
\maketitle

\section{Introduction}
\label{sec:introduction}

Sequential recommender systems play a critical role in digital platforms, including video streaming and e-commerce applications. 
These systems aim to predict the next item by analyzing sequential patterns in the user's historical interactions.
Existing methods utilize diverse neural architectures, including convolutional neural networks (CNNs)~\cite{caser,convformer}, recurrent neural networks (RNNs)~\cite{gru4rec,DBLP:conf/recsys/TanXL16}, and Transformer models~\cite{sasrec,bert4rec,ac-tsr}, to effectively model sequential user behaviors.
Conventional recommendation models predominantly employ a unique ID for item representation, capturing collaborative information in user interaction data.
While achieving remarkable success, these ID-based models neglect the rich textual metadata available in modern platforms, which is beneficial for modeling user preferences. 
In response to this issue, several recent studies~\cite{nova,dif-rec,dlfs-rec,kar,llmemb} have increasingly focused on harnessing textual information to enhance the sequential recommendation framework.

Early approaches~\cite{fdsa,unisrec} achieve moderate success by introducing text representations encoded by pre-trained language models (PLMs) in a straightforward way (\eg sum and concat), but these methods overlook the semantic gap between ID embeddings and text representations.
To address this limitation, some studies adopt various techniques, including contrastive learning~\cite{s3rec}, attention mechanisms~\cite{nova,dif-rec} and graph-based aggregation~\cite{mmsr}, to better fuse ID embeddings and text representations. 
Furthermore, recent studies propose mapping text representations into semantic codes for sequential recommendation~\cite{recjpq,vqrec,mmsr} or generative recommendation~\cite{tiger,lc-rec,actionpiece}.
These codes are typically derived through clustering or vector quantization based on text representations~\cite{howtoindex,tiger}.
A key merit of this approach is that the shared codes between different items reflect their semantic similarities.
Despite their notable effectiveness, these methods struggle to achieve an appropriate trade-off between the textual semantics embedded in text representations and the collaborative semantics implied by sequential patterns.
Integrating item ID embeddings with text representations may result in the model overemphasizing textual features while neglecting the collaborative relationships between items~\cite{vqrec}.
Although code-based methods can alleviate the over-reliance on text, they suffer from textual semantic loss in a certain extent.
% Additionally, code-based methods rely on code sharing relationships, which alleviates the over-reliance on textual features but inevitably leads to textual semantic loss.

Considering these issues, our idea is to employ semantic codes as a bridge to achieve a better balance between textual and collaborative information.
This approach is motivated by two key characteristics of semantic codes.
First, they can be regarded as a specialized form of item IDs, associated with learnable embeddings that can flexibly capture collaborative semantics by modeling interaction sequences.
Second, the multiple codes for each item imply semantics of different granularities or levels, injecting prior knowledge of semantic similarity into item representations, which can serve as guidance for refined textual-collaborative semantics integration.
To develop our approach, we focus on two aspects: (1) leveraging semantic codes as guidance to flexibly extract useful information from text representations, alleviating the over-reliance on textual information, and (2) learning informative item representations to enhance sequential recommendation.

To this end, we propose \textbf{CCFRec}, an innovative \underline{C}ode-based textual and \underline{C}ollaborative semantic \underline{F}usion approach for sequential \underline{Rec}ommendation.
Unlike previous methods that directly utilize text embeddings as item representations, our approach develops a code-guided fusion module to derive semantically enriched item representations. In this module, semantic codes serve as queries to flexibly extract relevant information from text representations, thereby mitigating the excessive dependence on item text while enabling a more comprehensive learning of code-based collaborative information.
To obtain semantic codes with different granularity information, we encode various item attributes (\eg title, brand, categories) separately as multi-view text representations and quantize them into a tuple of codes through vector quantization.
Then, we implement the semantic fusion module as a multi-layer cross-attention network to incorporate code embeddings and multi-view text representations.
In order to further enhance semantic fusion and sequence modeling, we propose two optimization objectives through item code masking, namely \emph{masked code modeling} and \emph{masked sequence alignment}.
For masked code modeling, the semantic fusion module recovers the masked code from multi-view text representations, which has two advantages. Firstly, it can capture the code relationships within individual items. Secondly, it promotes the integration between code embeddings and text representations.
For masked sequence alignment, we align the user preference representations derived from interaction sequences involving masked codes with the original one, aiming to enhance the robustness of user preference modeling.

In summary, the main contributions of this paper are as follows:

$\bullet$ We propose \textbf{CCFRec}, a novel code-based textual and collaborative semantic fusion method for sequential recommendation, which bridges the textual-collaborative gap through semantic codes.

$\bullet$ We design an optimization method via code masking that further enhances semantic fusion and sequence modeling through masked code modeling and masked sequence alignment.

$\bullet$ Extensive experiments on four public recommendation benchmarks demonstrate the effectiveness of our proposed framework CCFRec.

\begin{figure*}[]
    \centering
    \includegraphics[width=\linewidth]{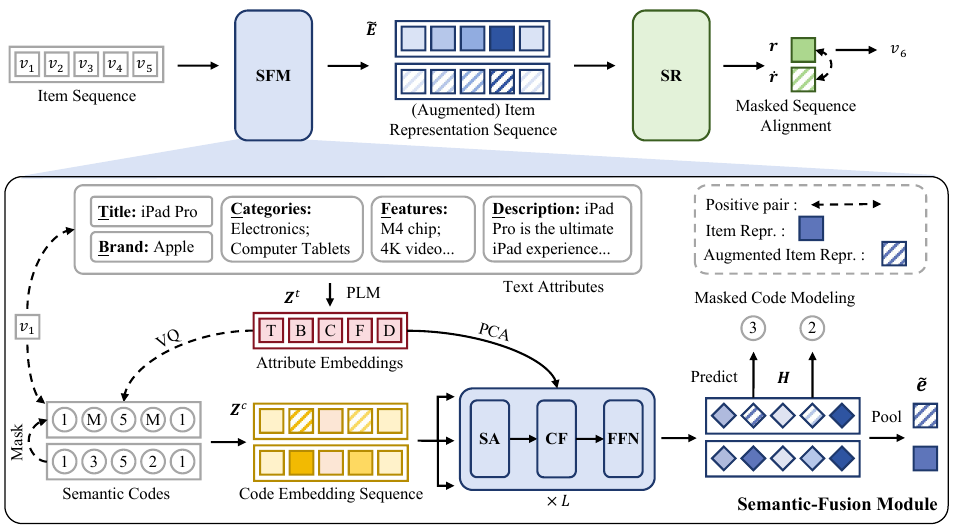}
    \captionsetup{font={small}}
    \caption{Overall framework of CCFRec, which comprises two key components, \ie the Semantic-Fusion Module (SFM) and the Sequential Recommendation backbone (SR). SFM stacks Self-Attention (SA) layers, Cross-Fusion (CF) layers, and Feed-Forward Networks (FFN) to enable effective learning of semantic-fused item representations.}
    \label{fig:model}
\end{figure*}

\section{Methodology}
\label{sec:methodology}

\subsection{Problem Statement}
In our scenario, we focus on sequential recommendation tasks. Let $\mathcal{U}=\{u_1,\dots,u_{|\mathcal{U}}|\}$ denote a set of users and $\mathcal{V}=\{v_1,\dots,v_{|\mathcal{V}|}\}$ denote a set of items. Each item $v$ is associated with a unique item ID and a series of text attributes $\mathcal{A}=\{a_1,\dots,a_m\}$ (\eg title, brand, category, \etc) where $m$ represents the attribute number. Here, each item attribute $a_i$ corresponds to a descriptive text. Additionally, $a_i$ is mapped to a tuple of semantic codes, denoted by $C_i=(c^i_1,\dots,c^i_k)$, where semantic codes under the same view (\eg title or brand) share a common code space. Given a user interaction sequence $s=\{v_1,v_2,\dots,v_t\}$ in chronological order, the objective of sequential recommendation is to predict the next item that the user is most likely to interact with, which is formulated as $\max P(v_{t+1}|\{v_1,\dots,v_t\})$.

\subsection{Text Embeddings and Semantic Codes}
In this part, we describe the process of deriving two distinct representations for items.
This involves (a) constructing multi-view text embeddings by encoding each item attribute individually, and (b) mapping the text embeddings to semantic codes through multi-level vector quantization methods. 

\subsubsection{Multi-View Text Embeddings}
Text embeddings are widely acknowledged for encapsulating rich textual information.
To effectively harness item content, previous studies~\cite{tedrec,llmemb,mmsr,kdsr} typically concatenate all associated text and encode it into a single embedding using pre-trained language models (PLMs). However, it's difficult to leverage fine-grained item information by jointly encoding all content. Moreover, due to the limited input length of PLMs, the item text often needs to be truncated, leading to potential incomplete semantics. Therefore, to fully exploit attribute-wise semantic information, we propose encoding each item attribute individually to obtain multi-view text embeddings. Formally, given text attributes $\mathcal{A}=\{a_{1},\dots,a_{m}\}$ of item $v$, the text embedding of attribute $a_{i}$ is formulated as
$$
    \bm{z}^t_{i}=\operatorname{PLM}(a_{i}),\ i\in\{1,\dots,m\},
$$
where $\bm{z}^t_{i}$ denotes the mean pooling representation of the output.

\subsubsection{Vector Quantization for Semantic Codes}
In our framework, semantic codes serve a dual purpose. First, they are linked to learnable embeddings, enabling the flexible capture of collaborative information. Second, they act as bridges to facilitate the fusion of textual and collaborative semantics. Unlike unique IDs, semantic codes allow semantically similar items to share representations, thereby capturing inherent relationships among items.
Moreover, the multi-level structure of these codes reflects textual semantics from different aspects, providing guidance for fine-grained semantic fusion. These characteristics make semantic codes particularly effective for item representation, as evidenced by prior research \cite{vqrec,tiger,recjpq}.
In our approach, we utilize vector quantization techniques, such as product quantization (PQ) and residual quantization (RQ), to map item attribute embeddings into multi-level discrete codes. Specifically, for an attribute embedding $\bm{z}^t_{i}$ of item $v$, the vector quantization process generates a tuple of discrete codes $C_i=(c^i_{1},\dots,c^i_{k})$, where $k$ denotes the number of quantized sub-vectors, which is a pre-defined hyper-parameter. Given that each item has $m$ distinct attributes, this process produces a total of $n_c=m\times k$ semantic codes, represented as $(c^1_1,c^1_2,\dots,c^m_{k-1},c^m_k)$.

\subsection{Code-Guided Semantic Fusion}
To effectively integrate textual and ID-based collaborative semantics, prior research~\cite{fdsa,s3rec,unisrec,mmsr,tedrec,dlfs-rec,smlp4rec} has typically combined text representations with ID embeddings or utilized the discrete codes derived from text embeddings to enhance sequential recommenders.
Despite their effectiveness, these methods often struggle to achieve an appropriate trade-off between the textual information encoded in text embeddings and the collaborative information implied by sequential patterns.
To address this issue, we propose to employ semantic codes as bridges to extract and integrate relevant information from text representations.
A key challenge in this approach lies in fully unleashing the potential of semantic codes to serve as effective bridges.
To tackle it, we introduce a code-guided semantic-fusion module that integrates text representations with multi-level semantic code embeddings using cross-attention mechanisms, where semantic codes are considered as the query and text representations are regarded as the key and value. This design enables the semantic codes to flexibly extract useful information from text representations, mitigating the over-reliance on item text.
Subsequently, a Transformer-based recommendation backbone leverages these fused item representations to model sequential user behavior and makes final recommendations.

\subsubsection{Semantic Code Embeddings}
Our approach leverages two types of embeddings, \ie attribute embeddings and code embeddings, to capture textual and collaborative information, respectively. Each item $v$ is associated with $n_c$ semantic codes denoted as $(c_{1}, \dots, c_{n_c})$ (for simplicity, we do not distinguish the attributes). To learn the collaborative correlations among items, we set up $n_c$ code embedding tables for lookup. For each position $l$, all items share a common code embedding table $\bm{E}^{(l)} \in \mathbb{R}^{C \times d}$, where $d$ denotes the dimension of the item representations. Using these tables, we look up the code embeddings for item $v$ as $\{\bm{z}^c_{i}\}_{i=1}^{n_c}$ where $\bm{z}^c_{i}\in\bm{E}^{(i)}$.
In addition to code embeddings, each item $v$ is also associated with $m$ attribute embeddings, denoted as $\{\bm{z}^t_{i}\}_{i=1}^m$. Here, $\bm{z}^t_{i} \in \mathbb{R}^{d_e}$ represents the embedding of the $i$-th attribute, where $d_e$ is typically larger than $d$. To ensure compatibility between two types of embeddings, we apply Principal Component Analysis (PCA) to reduce the dimension of $\bm{z}^t_{i}$ to $d$.

\subsubsection{Semantic-Fusion Module}
To obtain semantic-fused item representations, we propose a semantic-fusion module composed of multiple stacked blocks. Each block consists of a self-attention layer, a cross-fusion layer and a multilayer perceptron network.
Each item is associated with its attribute representations $\bm{Z}^t=[\bm{z}^t_{1},\dots,\bm{z}^t_{m}]\in\mathbb{R}^{m\times d}$ and code embeddings $\bm{Z}^c=[\bm{z}^c_{1},\dots,\bm{z}^c_{n_c}]\in\mathbb{R}^{n_c\times d}$. 
The code embeddings are first fed into the bi-directional self-attention layer to obtain mixed code representations. Then a cross-fusion layer based on the multi-head attention mechanism (denoted by MHA($Q,K,V$)) is employed to achieve the semantic fusion of the text and code sequences. The whole process can be formally written as:
\begin{align}
    \tilde{\bm{H}}^{l} &= \operatorname{MHA}\left(\bm{H}^{l-1},\bm{H}^{l-1},\bm{H}^{l-1}\right),\\
    \bm{H}^{l} &= \operatorname{FFN}\left(\operatorname{MHA}\left(\tilde{\bm{H}}^{l},\bm{Z}^t,\bm{Z}^t\right)\right),\ l\in\{1,\dots,L\},
\end{align}
where $\bm{H}^{0} = \bm{Z}^c$ and $L$ is the number of layers in the semantic-fusion module and $\operatorname{FFN}(\cdot)$ denotes the multilayer perceptron network.
The ordering of elements within these two sequences holds no intrinsic meaning. Therefore, we omit position embeddings to neglect sequential relationships.
The final semantically fused sequence representation, $\bm{H}\in\mathbb{R}^{n_c\times d}$, is aggregated to generate the semantic-fused item representation denoted as $\bm{e}=\operatorname{Pool}\left(\bm{H}\right)$, where $\operatorname{Pool}(\cdot)$ denotes mean pooling. To enhance semantic code embedding learning, the pooled semantic code representation is integrated, yielding the final item representation $\tilde{\bm{e}}=\bm{e}+\operatorname{Pool}\left(\bm{Z}^c\right)$.

\subsubsection{Recommendation Backbone}.
We adopt a widely used Transformer architecture as our recommendation backbone. Given an interaction sequence $s=\{v_1,v_2,\dots,v_n\}$, we can obtain the semantic-fused item representation matrix $\tilde{\bm{E}}=\{\tilde{\bm{e}}_1;\dots;\tilde{\bm{e}}_n\}\in\mathbb{R}^{n\times d}$ where [;] denotes the concatenation operation. The item representations $\tilde{\bm{e}}_i$ and absolute position embeddings $\bm{p}_i$ are summed up as the input to the backbone for sequential preference modeling.
Subsequently, the hidden states are updated through a multi-head attention (MHA) mechanism followed by a multilayer perceptron network (FFN).
This process is formally defined as:
\begin{align}
    \hat{\bm{h}}^0_i&=\tilde{\bm{e}}_i+\bm{p}_i,\ i\in\{1,\dots,n\},\\
    \hat{\bm{H}}^{l}&=\operatorname{FFN}\left(\operatorname{MHA}\left(\hat{\bm{H}}^{l-1},\hat{\bm{H}}^{l-1},\hat{\bm{H}}^{l-1}\right)\right),\ l\in\{1,\dots,L_r\},
\end{align}
where $\hat{\bm{H}}^l=\{\hat{\bm{h}}^l_1;\dots;\hat{\bm{h}}^l_n\}\in\mathbb{R}^{n\times d}$ denotes the hidden state sequence in the $l$-layer and $L_r$ denotes the total number of layers in the recommendation backbone.
The final hidden state corresponding to the last position is utilized as the representation of user preferences, denoted by $\bm{r}=\hat{\bm{H}}[-1]$.
Following prior studies~\cite{s3rec,fmlp-rec,tedrec}, we employ the cross-entropy loss as the training objective, which is written as
\begin{align}
\label{eq:ce}
    \mathcal{L}_\text{CE} = -\log\frac{\exp(g(\bm{r},\tilde{\bm{e}}_{n+1})/\tau)}{\sum_{j\in\mathcal{B}}\exp(g(\bm{r},\tilde{\bm{e}}_{j})/\tau)},
\end{align}
where $\mathcal{B}$ is the set of sampled negative items, $g(\cdot,\cdot)$ denotes the cosine similarity measure and $\tau$ is a temperature coefficient.

\subsection{Enhanced Representation Learning via Code Masking}
Compared with existing semantic fusion methods~\cite{fdsa,mmsr,tedrec}, optimizing our framework is more challenging due to its two hierarchically arranged components. A single recommendation optimization objective is insufficient to ensure the comprehensive learning of the entire framework, particularly for the code representations.
To address this issue, we propose two optimization objectives based on code masking to further enhance the representation learning in CCFRec. Specifically, we introduce the Masked Code Modeling (MCM) task, which promotes the semantic fusion between text and codes to refine item representations. Additionally, we incorporate the Masked Sequence Alignment (MSA) task to improve the sequence modeling capabilities of the recommendation backbone.

\paratitle{Masked Code Modeling}.
To foster the semantic fusion between text representations and code embeddings, we present the Masked Code Modeling (MCM) task, inspired by the success of Masked language modeling (MLM) in natural language processing.
MLM is widely used to predict randomly masked word tokens in sequences by conditioning on the unmasked tokens. In MCM, at each training step, a random $\rho$ percent of the codes in the semantic code sequence is masked following the same strategy described in BERT~\cite{bert}.
The semantic-fusion module is then tasked with recovering the masked codes based on attribute embeddings and the remaining unmasked codes.
Formally, given a semantic code sequence $s^c=\{c_1,\dots,c_{n_c}\}$ of item $v$, we randomly mask a $\rho$ ratio of the codes and obtain the masked sequence $\dot{s}^c=\{c_1,\dots,[MASK],\dots,c_{n_c}\}$ and let $\mathcal{M}^c$ represent the set of mask codes. Then, the code sequence representation produced by the semantic-fusion module is denoted as $\{\dot{\bm{h}}_i\}^n_{i=1}$ where $n$ represents the sequence length.
The MCM loss is formulated as 
\begin{align}
    \label{eq:mcm}
    \mathcal{L}_{\text{MCM}}=\frac{1}{|\mathcal{M}^c|}\sum_{x\in\mathcal{M}^c}-\log\frac{\exp(g(\dot{\bm{h}}_x,\bm{z}^c_{x})/\tau)}{\sum_{j\in\mathcal{C}}\exp(g(\dot{\bm{h}}_x,\bm{z}^c_{j})/\tau)},
\end{align}
where $\mathcal{C}$ denotes the set of all semantic codes. This approach offers two main advantages: first, it strengthens the correlations among semantic codes; second, it improves the alignment between text and code embeddings through text-dependent masked code reconstruction, thereby facilitating more effective textual-collaborative semantic fusion.

\paratitle{Masked Sequence Alignment}.
MCM is primarily used to optimize the SFM. In order to enhance the sequence representation learning of the recommendation backbone, we introduce the Masked Sequence Alignment (MSA) task, using the contrastive learning technique that has been widely validated in sequential recommendation tasks~\cite{cl4srec,iclrec,duorec,maerec}.
In MSA, code masking serves as an effective data augmentation strategy to generate augmented item representations. This approach increases the diversity of inputs to the recommendation backbone, thereby refining its ability to model user sequences.
Specifically, we first generate augmented item representations by feeding the masked code sequence and associated text representations into the SFM. These augmented item representations are then passed to the recommendation backbone to produce augmented sequence representations. Subsequently, the two sequence representations are aligned using the InfoNCE loss.
The representations derived from the same sequence are considered as positive samples, while those from different sequences within the same batch are treated as negative samples.
The objective function for \textit{masked sequence alignment} is formulated as follows:
\begin{align}
\label{eq:msa}
\mathcal{L}_{\text{MSA}} = \frac{1}{2}\Biggl(
\operatorname{InfoNCE}(\dot{\bm{r}},\bm{r},\mathcal{B}_{\bm{r}}) + \operatorname{InfoNCE}(\bm{r},\dot{\bm{r}},\mathcal{B}_{\dot{\bm{r}}})
\Biggr),
\end{align}
where $\mathcal{B}_{\bm{r}}$ and $\mathcal{B}_{\dot{\bm{r}}}$ denote the batch of original sequence representations and augmented sequence representations generated by the recommendation backbone, respectively. $\operatorname{InfoNCE}(\cdot,\cdot,\cdot)$ denotes the InfoNCE loss, which can be written as:
\begin{align}
    \operatorname{InfoNCE}(\bm{x},\bm{y}^+,\mathcal{R}_{\bm{y}})= -\log\frac{\exp(g(\bm{x},\bm{y}^+)/\tau)}{\sum_{\bm{y}\in\mathcal{R}_{\bm{y}}}\exp(g(\bm{x},\bm{y})/\tau)},
\end{align}
where $\bm{x}$ and $\bm{y}^+$ represent a pair of positive instances and $\mathcal{R}_{\bm{y}}$ denotes a set of samples including both positive and negative instances.

Finally, the overall optimization objectives can be denoted as follows:
\begin{align}
    \mathcal{L}=\mathcal{L}_\text{CE}+\alpha\mathcal{L}_\text{MCM}+\beta\mathcal{L}_\text{MSA},
\end{align}
where $\alpha$ and $\beta$ are hyper-parameters that control the trade-off between the respective training objectives. 

\subsection{Discussion}

To highlight the novelty and contributions of our approach, we compare CCFRec with existing text-enhanced methods, as shown in Table~\ref{tab:discussion}.
Firstly, previous methods typically adopt text representations and ID embeddings to capture textual and collaborative information, respectively.
As a result, these two types of representations remain isolated and lack information sharing prior to their interaction (\eg sum or attention mechanisms).
In contrast, \textbf{\emph{we establish a connection between them by adopting semantic codes to capture collaborative information}}. The connection between semantic codes and text representations allows codes to serve as bridges connecting the two types of information.
Secondly, we adopt the \textbf{\emph{code-guided cross attention}} mechanism for semantic fusion by considering semantic codes as queries to extract useful information from text representations, enabling flexible textual semantic integration.
Thirdly, we introduce the masked code modeling task to capture the \textbf{\emph{code relationships within individual items}} and improve the fusion between text representations and code embeddings.

\begin{table}[]
\centering
    \captionsetup{font={small}}
    \caption{Comparison of different methods. `Flexible' denotes the flexible integration of textual information. `CRM' denotes the code relation modeling within individual items.}
    \label{tab:discussion}
    \resizebox{\columnwidth}{!}{
    \renewcommand\arraystretch{1.0}
    \begin{tabular}{lcccc}
    \toprule
    Methods	& Input & Fusion Type & Flexible & CRM   \\
    \midrule
    UniSRec~\cite{unisrec} & Text \& ID &	Vector addition & \scalebox{0.8}{\textcolor{purple}{\XSolidBrush}} & \scalebox{0.8}{\textcolor{purple}{\XSolidBrush}}\\
    VQRec~\cite{vqrec}    & Code &	Vector quantization &  \scalebox{0.8}{\textcolor{teal}{\CheckmarkBold}} & \scalebox{0.8}{\textcolor{purple}{\XSolidBrush}}\\ 
    TedRec~\cite{tedrec}  & Text \& ID &	Contextual convolution & \scalebox{0.8}{\textcolor{purple}{\XSolidBrush}} & \scalebox{0.8}{\textcolor{purple}{\XSolidBrush}}\\
    CCFRec~(ours)	&	Text \& Code &	Code-guided attention & \scalebox{0.8}{\textcolor{teal}{\CheckmarkBold}} & \scalebox{0.8}{\textcolor{teal}{\CheckmarkBold}}\\
    \bottomrule
\end{tabular}
}
\end{table}

\section{Time Complexity}

\paratitle{Training}.
As outlined in Section~\ref{sec:methodology}, CCFRec consists of two key components, the semantic-fusion module (SFM) and the sequential recommendation backbone (SR). For the SFM, the computational complexities of the self-attention layer, the cross-fusion layer, and the feed-forward network when processing a single item are $\mathcal{O}(M^2d)$, $\mathcal{O}(MKd)$ and $\mathcal{O}(Md^2)$, respectively. Here, $d$ denotes the model dimension, while $M$ and $K$ represent the length of the semantic code sequence and item attribute sequence.
Consequently, the overall time complexity for deriving a semantic-fused item representation is $\mathcal{O}(M^2d+MKd+Md^2)$. For the SR, which utilizes a SASRec-like architecture, the time consumption of user behavior modeling is $\mathcal{O}(N^2d+Nd^2)$, where $N$ is the length of the user interactions. Thus, the total training complexity of CCFRec is $\mathcal{O}(N^2d+Nd^2+NM^2d+NMKd+NMd^2)$.
For brevity, the computation of training objectives is omitted.
Notably, in practical scenarios, the code length $M$ and attribute number $K$ are typically much smaller than interaction sequence length $N$ (\ie $M, K \ll N$). This ensures that the additional computational overhead introduced by CCFRec remains acceptable for real-world applications.

\paratitle{Deployment}.
Our proposed framework can be efficiently deployed in a lightweight manner by eliminating the SFM module. During deployment, as the learned representations of text attributes and semantic codes remain consistent for each item, we can precompute and store their semantic-fused representations by processing the attribute and code sequences through the SFM in advance. Consequently, the inference complexity is maintained at $\mathcal{O}(N^2d+Nd^2)$, which is identical to that of the mainstream sequential recommendation models such as SASRec.
\section{Experiments}
\label{sec:experiments}
In this section, we empirically demonstrate the effectiveness of our proposed framework CCFRec through extensive experiments and rigorous analysis.

\subsection{Experiment Setup}

\subsubsection{Dataset}
We conduct experiments on four subsets of the latest Amazon 2023 review data~\cite{amazon2023}, \ie ``\emph{Musical Instruments}'', ``\emph{Video Games}'', ``\emph{Baby Products}'' and ``\emph{Industrial Scientific}''.
To ensure a robust evaluation, we follow the same preprocessing steps (\ie 5-core filtering) as described in prior studies~\cite{s3rec,fmlp-rec,tedrec} to remove inactive users and unpopular items with less than five interactions.
Next, we organize user behavior sequences in chronological order, limiting the maximum sequence length to 20 items. We categorize item attributes into five fields: \emph{title}, \emph{brand}, \emph{categories}, \emph{features}, and \emph{description}, to capture fine-grained and rich semantic information. Each field's content is truncated to a maximum of 512 tokens per item.
The detailed statistics of the preprocessed datasets are presented in Table~\ref{tab:data_statistics}.

\subsubsection{Baseline Models}
We compare the proposed CCFRec with various sequential recommendation (SR) baselines, including the following three categories:
\noindent (1) \emph{Traditional SR models}:
{\textbf{GRU4Rec}}~\cite{gru4rec} adopts GRUs to capture sequential patterns in user interactions.
{\textbf{BERT4Rec}}~\cite{bert4rec} applies the bidirectional self-attention mechanism and mask prediction task for sequential recommendation.  
{\textbf{SASRec}}~\cite{sasrec} utilizes a unidirectional self-attentive model to predict the next item of interest.
{\textbf{FMLP-Rec}}~\cite{fmlp-rec} introduces an all-MLP model with learnable filters to reduce the noise in sequence modeling.
\noindent (2) \emph{CL-based SR models}:
{\textbf{S$^3$-Rec}}~\cite{s3rec} employs the correlation between items and features as self-supervised signals to empower user behavior modeling.
{\textbf{DuoRec}}~\cite{duorec} proposes a model-level data augmentation approach based on Dropout to resolve the representation degeneration.
{\textbf{MAERec}}~\cite{maerec} leverages a graph-masked autoencoder to adaptively enhance the sequential recommender.
\noindent (3) \emph{Text-enhanced SR models}:
{\textbf{FDSA}}~\cite{fdsa}  captures the item-level and feature-level correlations through a dual-stream self-attentive network.
{\textbf{UniSRec}}~\cite{unisrec} learns transferable textual item representations through a MoE-enhanced adaptor. We implement two variants of it: (i) UniSRec$_\text{T}$ with only text representations, and (ii) UniSRec$_\text{ID+T}$ with both text and ID representations. 
{\textbf{VQRec}}~\cite{vqrec} introduces vector-quantized item representations for sequential recommendation.
{\textbf{MMSR}}~\cite{mmsr} proposes a graph-based method to integrate multi-modal information in an adaptive order for enhanced sequential recommenders.
{\textbf{TedRec}}~\cite{tedrec} achieves sequence-level text-ID semantic fusion in the frequency domain, improving sequential recommendation performance.

\begin{table}[]
    \centering
    \caption{Statistics of the preprocessed datasets. Avg.\textnormal{\textit{len}} denotes the average length of the historical interaction sequences.}
    \resizebox{\columnwidth}{!}{
    \begin{tabular}{lrrrrr}
    \toprule
     Dataset    &\#Users   &\#Items   &\#Interactions &Sparsity &Avg.\textit{len} \\
     \midrule
     Instrument &57,439  &24,587  &511,836  &99.964\% &8.91 \\
     Scientific &50,985  &25,848  &412,947  &99.969\% &8.10\\
     Game &94,762  &25,612  &814,586  &99.966\% &8.60\\
     Baby &150,777  &36,013  &1,241,083  &99.977\% &8.23\\
     \bottomrule
    \end{tabular}}
    \label{tab:data_statistics}
\end{table}

\begin{table*}[]
\centering
\captionsetup{font={small}}
\caption{Performance comparisons among different methods. The best and second-best results are highlighted in bold and underlined font, respectively.
``N@K'' and ``R@K'' represent abbreviations of ``NDCG@K'' and ``Recall@K'', respectively. ``PQ'' and ``RQ'' are short for ``Product Quantization'' and ``Residual Quantization''.
``Improv.'' denotes the relative improvement ratios of CCFRec over the top-performing baselines.}
\label{tab:main_result}
\resizebox{\textwidth}{!}{%
\renewcommand\arraystretch{1.1}
\setlength{\tabcolsep}{1mm}{
\begin{tabular}{lrrrrrrrrrrrrrrrr}
\toprule
\multicolumn{1}{l}{\multirow{2}{*}{Methods}} & \multicolumn{4}{c}{Instrument} & \multicolumn{4}{c}{Scientific} & \multicolumn{4}{c}{Game} & \multicolumn{4}{c}{Baby} \\
\cmidrule(l){2-5} \cmidrule(l){6-9} \cmidrule(l){10-13} \cmidrule(l){14-17}
& \multicolumn{1}{c}{R@5} & \multicolumn{1}{c}{R@10} & \multicolumn{1}{c}{N@5} & \multicolumn{1}{c}{N@10} & \multicolumn{1}{c}{R@5} & \multicolumn{1}{c}{R@10} & \multicolumn{1}{c}{N@5} & \multicolumn{1}{c}{N@10} & \multicolumn{1}{c}{R@5} & \multicolumn{1}{c}{R@10} & \multicolumn{1}{c}{N@5} & \multicolumn{1}{c}{N@10} & \multicolumn{1}{c}{R@5} & \multicolumn{1}{c}{R@10} & \multicolumn{1}{c}{N@5} & \multicolumn{1}{c}{N@10} \\ \midrule \midrule
GRU4Rec & 0.0339 & 0.0540 & 0.0216 & 0.0281 & 0.0230 & 0.0374 & 0.0148 & 0.0194 & 0.0530 & 0.0820 & 0.0350 & 0.0443 & 0.0219 & 0.0354 & 0.0140 & 0.0184 \\
BERT4Rec & 0.0307 & 0.0485 & 0.0195 & 0.0252 & 0.0186 & 0.0296 & 0.0119 & 0.0155 & 0.0460 & 0.0735 & 0.0298 & 0.0386 & 0.0181 & 0.0300 & 0.0116 & 0.0154 \\
SASRec & 0.0333 & 0.0523 & 0.0213 & 0.0274 & 0.0259 & 0.0412 & 0.0150 & 0.0199 & 0.0535 & 0.0847 & 0.0331 & 0.0438 & 0.0221 & 0.0362 & 0.0135 & 0.0180 \\
FMLP-Rec & 0.0339 & 0.0536 & 0.0218 & 0.0282 & 0.0269 & 0.0422 & 0.0155 & 0.0204 & 0.0528 & 0.0857 & 0.0338 & 0.0444 & 0.0228 & 0.0367 & 0.0146 & 0.0190 \\ \midrule
S$^3$Rec & 0.0317 & 0.0496 & 0.0199 & 0.0257 & 0.0263 & 0.0418 & 0.0171 & 0.0219 & 0.0485 & 0.0769 & 0.0315 & 0.0406 & 0.0216 & 0.0338 & 0.0126 & 0.0168 \\
DuoRec & 0.0373 & 0.0573 & 0.0246 & 0.0310 & 0.0245 & 0.0379 & 0.0166 & 0.0209 & 0.0559 & 0.0844 & 0.0378 & 0.0469 & 0.0211 & 0.0346 & 0.0139 & 0.0182 \\
MAERec & 0.0369 & 0.0577 & 0.0243 & 0.0310 & 0.0303 & 0.0452 & 0.0205 & 0.0253 & 0.0618 & 0.0936 & 0.0411 & 0.0513 & 0.0224 & 0.0370 & 0.0146 & 0.0193 \\ \midrule
FDSA & 0.0369 & 0.0576 & 0.0240 & 0.0307 & 0.0273 & 0.0416 & 0.0183 & 0.0229 & 0.0544 & 0.0852 & 0.0361 & 0.0460 & 0.0243 & 0.0387 & 0.0158 & 0.0205 \\
UniSRec$_{\text{T}}$ & 0.0376 & 0.0589 & 0.0244 & 0.0312 & 0.0296 & 0.0469 & 0.0191 & 0.0246 & 0.0587 & 0.0925 & 0.0372 & 0.0480 & 0.0239 & 0.0380 & 0.0153 & 0.0198 \\
UniSRec$_{\text{ID+T}}$ & 0.0370 & 0.0598 & 0.0234 & 0.0308 & 0.0286 & 0.0457 & 0.0157 & 0.0214 & 0.0563 & 0.0921 & 0.0347 & 0.0459 & 0.0229 & 0.0386 & 0.0140 & 0.0190 \\
VQRec & 0.0379 & 0.0602 & 0.0227 & 0.0298 & 0.0293 & 0.0461 & 0.0170 & 0.0224 & 0.0581 & 0.0926 & 0.0355 & 0.0466 & 0.0260 & 0.0409 & 0.0166 & 0.0214 \\
MMSR & 0.0360 & 0.0569 & 0.0231 & 0.0300 & 0.0264 & 0.0427 & 0.0170 & 0.0208 & 0.0558 & 0.0881 & 0.0372 & 0.0461 & 0.0232 & 0.0389 & 0.0142 & 0.0185 \\
TedRec & 0.0374 & 0.0570 & 0.0249 & 0.0313 & 0.0282 & 0.0415 & 0.0195 & 0.0238 & 0.0624 & 0.0956 & \textbf{0.0419} & {\ul 0.0526} & 0.0246 & 0.0380 & 0.0165 & 0.0208 \\ \midrule
CCFRec(RQ) & {\ul 0.0426} & {\ul 0.0670} & {\ul 0.0273} & {\ul 0.0351} & {\ul 0.0348} & {\ul 0.0540} & {\ul 0.0214} & {\ul 0.0276} & {\ul 0.0642} & {\ul 0.1023} & 0.0400 & 0.0523 & {\ul 0.0285} & {\ul 0.0443} & {\ul 0.0182} & {\ul 0.0232} \\
CCFRec(PQ) & \textbf{0.0432} & \textbf{0.0682} & \textbf{0.0281} & \textbf{0.0361} & \textbf{0.0364} & \textbf{0.0555} & \textbf{0.0224} & \textbf{0.0285} & \textbf{0.0658} & \textbf{0.1042} & {\ul 0.0413} & \textbf{0.0536} & \textbf{0.0286} & \textbf{0.0455} & \textbf{0.0184} & \textbf{0.0238} \\
Improv. & +14.0\% & +13.3\% & +12.9\% & +15.3\% & +20.1\% & +18.3\% & +9.3\% & +12.7\% & +5.5\% & +9.0\% & -- & +1.9\% & +10.0\% & +11.3\% & +10.8\% & +11.2\% \\ \bottomrule
\end{tabular}%
}}
\end{table*}

\subsubsection{Evaluation Settings}
To evaluate the performance of sequential recommendation task, we adopt top-$K$ Recall and top-$K$ Normalized Discounted Cumulative Gain (NDCG) as the evaluation metrics, where $K$ is set to 5 and 10. 
In line with prior studies~\cite{s3rec,tiger}, we employ the \emph{leave-one-out} strategy for dataset splitting.
For each user interaction sequence, the last interacted item is used as testing data, the second-last item is used as validation data, and all remaining items are used for training.
We compute the ranking results across the entire item set to guarantee rigorous evaluations. 

\subsubsection{Implementation Details}
For CCFRec, we leverage Sentence-T5~\cite{sentence-t5} to encode the text attributes associated with each item as its text embeddings. Both the semantic-fusion module and the sequential recommender are configured with 2 layers and 2 attention heads, with a hidden dimension of 512. The embedding size is fixed at 128, while each attribute vector is quantized into 4 sub-vectors (\ie $k=4$) with a codebook size of 256. The temperature coefficient $\tau$ and mask ratio $\rho$ are set to 0.07 and 0.5, respectively. We adopt the Adam optimizer for model training, with the learning rate tuned in \{0.003, 0.001, 0.0005\}. The dropout probability is tuned within \{0.1, 0.2, 0.3, 0.4, 0.5\} and the loss coefficients $\alpha$ and $\beta$ are searched in \{0.2, 0.4, 0.6, 0.8\} for optimal performance. To avoid overfitting, we utilize an early-stopping strategy, halting training when NDCG@10 in the validation set shows no improvement over 10 consecutive epochs.
We implement the majority of baseline models using the open-source recommendation system library RecBole~\cite{recbole,recbole2.0,recbole-new}. For CL-based SR models, namely DuoRec and MAERec, we utilize the self-supervised learning framework SSLRec~\cite{sslrec}. For fair comparisons, we fix the embedding size of all models to 128 and perform the hyperparameter grid search for optimal results. Furthermore, all Transformer-based baseline models adopt the same recommendation backbone architecture as our CCFRec for fairness.

\begin{table*}[]
\centering
\captionsetup{font={small}}
\caption{Ablation study of our proposed method on three datasets. The best and second-best results are denoted in bold and underlined fonts, respectively.}
\label{tab:ablation}
\resizebox{0.98\textwidth}{!}{%
\renewcommand\arraystretch{0.95}
\begin{tabular}{lrrrrrrrrrrrr}
\toprule
\multirow{2}{*}{Variants} & \multicolumn{4}{c}{Instrument} & \multicolumn{4}{c}{Scientific} & \multicolumn{4}{c}{Baby} \\
\cmidrule(l){2-5} \cmidrule(l){6-9} \cmidrule(l){10-13}
 & \multicolumn{1}{c}{R@5} & \multicolumn{1}{c}{R@10} & \multicolumn{1}{c}{N@5} & \multicolumn{1}{c}{N@10} & \multicolumn{1}{c}{R@5} & \multicolumn{1}{c}{R@10} & \multicolumn{1}{c}{N@5} & \multicolumn{1}{c}{N@10} & \multicolumn{1}{c}{R@5} & \multicolumn{1}{c}{R@10} & \multicolumn{1}{c}{N@5} & \multicolumn{1}{c}{N@10} \\ \midrule
(0) CCFRec & \textbf{0.0432} & \textbf{0.0682} & \textbf{0.0281} & \textbf{0.0361} & \textbf{0.0364} & \textbf{0.0555} & \textbf{0.0224} & \textbf{0.0285} & \textbf{0.0286} & \textbf{0.0455} & \textbf{0.0184} & \textbf{0.0238} \\
(1) \  \emph{w/o} $\mathcal{L}_{\text{MCM}}$ & 0.0411 & {0.0668} & 0.0260 & {0.0343} & 0.0348 & {\ul 0.0548} & 0.0214 & {\ul 0.0279} & 0.0275 & {\ul 0.0444} & 0.0176 & 0.0230 \\
(2) \  \emph{w/o} $\mathcal{L}_{\text{MSA}}$ & 0.0414 & 0.0664 & 0.0262 & 0.0342 & {\ul 0.0356} & {\ul 0.0548} & {\ul 0.0215} & 0.0277 & 0.0275 & 0.0439 & 0.0176 & 0.0229 \\
(3) \  \emph{w/o} Text Emb & {0.0416} & 0.0649 & {0.0268} & {0.0343} & 0.0343 & 0.0538 & 0.0212 & 0.0274 & {\ul 0.0281} & {\ul 0.0444} & {\ul 0.0183} & {\ul 0.0236} \\
(4) \  Random Code & 0.0406 & 0.0641 & 0.0256 & 0.0331 & 0.0317 & 0.0501 & 0.0191 & 0.0250 & 0.0270 & 0.0427 & 0.0171 & 0.0222 \\
(5) \  Global Emb & {\ul 0.0422} & {\ul 0.0671} & {\ul 0.0273} & {\ul 0.0352} & 0.0330 & 0.0533 & 0.0203 & 0.0268 & 0.0278 & 0.0438 & 0.0178 & 0.0229 \\
(6) \  \emph{w/o} CA & 0.0408 & 0.0640 & 0.0251 & 0.0326 & 0.0319 & 0.0509 & 0.0185 & 0.0246 & 0.0262 & 0.0427 & 0.0167 & 0.0220 \\
\bottomrule
\end{tabular}%
}
\end{table*}

\subsection{Overall Performance}
We compare CCFRec with various baseline approaches on four public benchmarks and present the overall results in Table~\ref{tab:main_result}.
From the results, we have the following findings:

For traditional SR models, FMLP-Rec demonstrates superior performance than SASRec by replacing the self-attention layer with filter-enhanced MLPs. CL-based SR models (\ie S$^3$Rec, DuoRec, MAERec) generally outperform traditional SR models (\ie GRU4Rec, BERT4Rec, SASRec, FMLP-Rec), highlighting the effectiveness of contrastive learning techniques in enhancing sequence representation learning.
Notably, the pure ID-based MAERec exhibits better performance than most text-enhanced SR models (\ie FDSA, UniSRec, VQRec, MMSR) in the Scientific and Game datasets, underscoring the critical role of collaborative information in modeling user interactions.

Regarding text-enhanced SR models, they typically outperform other SR models by leveraging auxiliary textual information from item contents.
UniSRec$_\text{T}$ performs better than UniSRec$_\text{ID+T}$ indicating that the fusion of textual information and ID-based collaborative information remains suboptimal. VQRec achieves outstanding results in the Baby, demonstrating the effectiveness of vector-quantized item representations. TedRec excels by achieving sequence-level semantic fusion between text and IDs in the frequency domain, showing the best performance across most datasets.

Finally, our proposed CCFRec(PQ) achieves the best or second-best performance across all datasets, significantly outperforming the best baseline models in most cases. CCFRec(RQ) also achieves competitive results but performs slightly worse than CCFRec(PQ), which may be attributed to the fact that PQ-based semantic codes preserving semantics of different levels are more suitable for our framework than RQ-based codes implying semantics of different granularities.
Different from existing text-enhanced methods, we propose leveraging semantic codes to bridge the gap between textual and collaborative information. With the specially designed semantic-fusion module and optimization objectives, including masked code modeling and masked sequence alignment, we achieve effective textual-collaborative semantic fusion for sequential recommendations, significantly improving the model performance.

\subsection{Ablation Study}
\label{sec:ablation}
To evaluate the contribution of each proposed component, we conduct ablation studies on the Instrument, Scientific, and Baby datasets. Table~\ref{tab:ablation} presents the comparative results of the following six variants:
(1) \underline{\emph{w/o} $\mathcal{L}_{\text{MCM}}$} removes the masked code modeling (MCM) defined in Eq.~\eqref{eq:mcm}.
(2) \underline{\emph{w/o} $\mathcal{L}_{\text{MSA}}$} without the masked sequence alignment (MSA) in Eq.~\eqref{eq:msa}.
(3) \underline{\emph{w/o} Text Emb} substitutes text embeddings with semantic codes as cross-fusion layer input.
(4) \underline{Random Code} replaces the semantic codes derived from text embeddings with random codes.
(5) \underline{Global Emb} encodes all text attributes into a single global embedding, replacing the individual attribute embeddings
(6) \underline{\emph{w/o} CA} replaces the SFM with mean pooling of all associated text representations and code embeddings to generate the final item representation.

The experimental results reveal that removing any component from CCFRec consistently degrades model performance.
This performance decline underscores the critical role of both proposed optimization objectives (i.e., MCM and MSA) in improving the representation learning of the overall framework.
Furthermore, it demonstrates the significance of textual and collaborative information, as captured by text representations and code embeddings, respectively.

\subsection{Further Analysis}

\begin{table}[]
\centering
\captionsetup{font={small}}
\caption{Ablation studies of item attributes on Instrument and Scientific datasets.}
\resizebox{\columnwidth}{!}{
\renewcommand\arraystretch{1.0}
\begin{tabular}{ccccc|cc|cc}
\hline
\multirow{2}{*}{Title} & \multirow{2}{*}{Brand} & \multirow{2}{*}{Feat.} & \multirow{2}{*}{Cate.} & \multirow{2}{*}{Desc.} & \multicolumn{2}{c|}{{Instrument}} & \multicolumn{2}{c}{{Scientific}} \\
 \cline{6-9} & & & & &R@10 & N@10  & R@10 & N@10 \\ \hline
\usym{2713} &             &             &      &  
& 0.0610 & 0.0319 & 0.0520 & 0.0262\\ 
\usym{2713} & \usym{2713} &             &      &  
 & 0.0628 & 0.0329 & 0.0531 & 0.0267 \\ 
\usym{2713} & \usym{2713} & \usym{2713} &      &  
 & 0.0653 & 0.0344 & 0.0545 & 0.0278\\
\usym{2713} & \usym{2713} & \usym{2713} & \usym{2713} &  
 & 0.0667 & 0.0355 & 0.0552 & 0.0283\\
\usym{2713} & \usym{2713} & \usym{2713} & \usym{2713} & \usym{2713} 
 & \textbf{0.0682} & \textbf{0.0361} & \textbf{0.0555} & \textbf{0.0285}\\
 \hline
\end{tabular}}

\label{tab:attr_ablation}
\end{table}

\subsubsection{Impact of Item Attributes}
To investigate how each item attribute affects the final performance, we conduct ablation studies on the Instrument and Scientific datasets by sequentially adding individual attributes. The results are reported in Table~\ref{tab:attr_ablation}, from which we can find that text contents have a significant impact on the final performance. In general, richer attribute contents enable CCFRec to extract more textual information beneficial for recommendation, allowing more refined item representations and user preference modeling.

\subsubsection{Impact of Code Number}

\begin{table}[]
\centering
\captionsetup{font={small}}
\caption{Performance analysis of code number in CCFRec. $k$, $m$ and $C$ denote the number of sub-vectors in vector quantization, the number of attributes and the codebook size of each sub-vector, respectively.}
\label{tab:code_num}
\resizebox{\columnwidth}{!}{%
\begin{tabular}{lrrrrrrrr}
\toprule
\multirow{1}{*}{Number} & \multicolumn{4}{c}{Instrument} & \multicolumn{4}{c}{Scientific} \\
\cmidrule(l){2-5} \cmidrule(l){6-9}
\multirow{1}{*}{($k\times m \times C$)} & \multicolumn{1}{c}{R@5} & \multicolumn{1}{c}{R@10} & \multicolumn{1}{c}{N@5} & \multicolumn{1}{c}{N@10} & \multicolumn{1}{c}{R@5} & \multicolumn{1}{c}{R@10} & \multicolumn{1}{c}{N@5} & \multicolumn{1}{c}{N@10} \\ \midrule
$2\times5\times256$ & 0.0400 & 0.0649 & 0.0256 & 0.0337 & 0.0338 & 0.0542 & 0.0208 & 0.0274 \\
$4\times5\times256$ & 0.0432 & 0.0682 & 0.0281 & 0.0361 & 0.0364 & 0.0555 & 0.0224 & 0.0285 \\
$8\times5\times256$ & \textbf{0.0445} & \textbf{0.0696} & \textbf{0.0285} & \textbf{0.0368} & \textbf{0.0375} & \textbf{0.0564} & \textbf{0.0232} & \textbf{0.0288} \\
\bottomrule
\end{tabular}%
}
\end{table}

We analyze the impact of code number on the final recommendation performance of CCFRec by changing the number of sub-vectors in the vector quantization from 2 to 8. The results in Table~\ref{tab:code_num} indicate that CCFRec consistently benefits from increasing the code number.
The reason is that a larger code number allows CCFRec to capture textual semantics and model the item correlations more effectively in a broader representation space.
However, increasing the number of codes also raises computational resource requirements. Therefore, it is essential to strike a balance between model performance and computational efficiency.

\subsubsection{Impact of Code Type}

\begin{figure}[]
    \centering
    \includegraphics[width=0.97\linewidth]{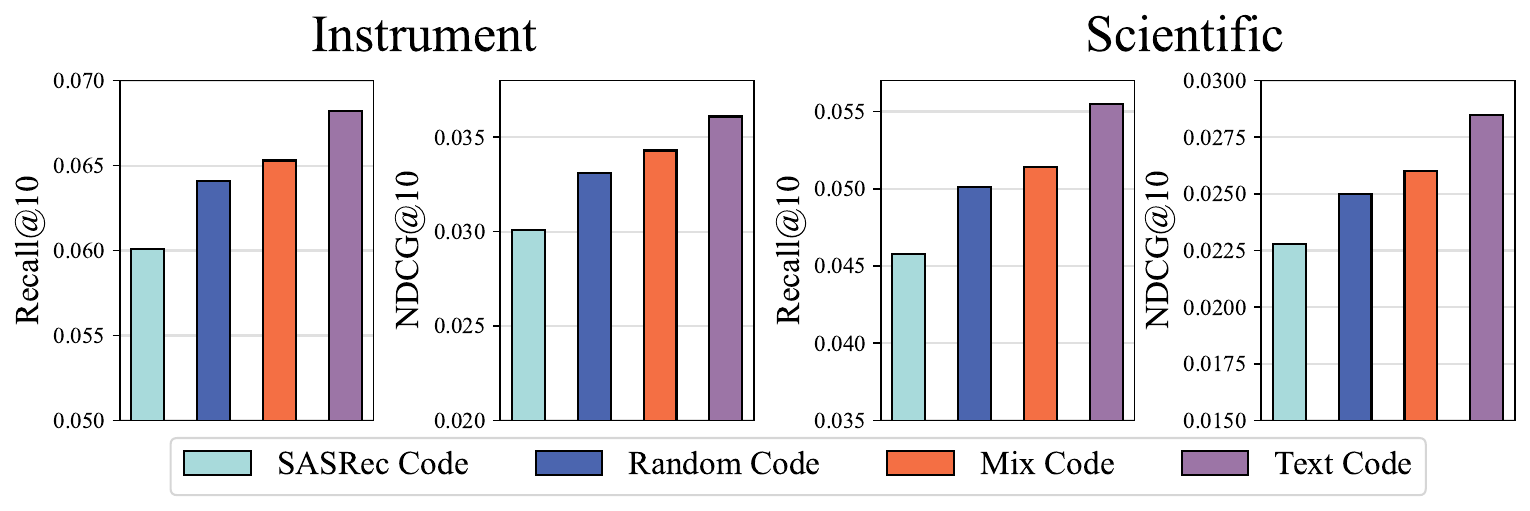}
    \captionsetup{font={small}}
    \caption{Performance comparison of different semantic codes. `Mix Code' refers to a combination of half text-based codes and half SASRec-based codes.}
    \label{fig:code_type}
\end{figure}

To examine the impact of code types, we replace the original text-based semantic codes with three alternatives: random codes, codes derived from SASRec embeddings and mixed codes that combine equal proportions of text-based and SASRec-based codes. To ensure a fair comparison, the number of semantic codes in each case is kept consistent.
As shown in Figure~\ref{fig:code_type}, the use of SASRec-based semantic codes leads to significant performance degradation. This is primarily because the collaborative information captured by SASRec-based codes is redundant, as CCFRec is already capable of learning it.
Our results indicate that models perform better when using more text semantic codes (Text > Mix > SASRec), underscoring the importance of textual semantic codes.
Interestingly, random codes outperform SASRec-based codes, likely due to their stronger regularization effect.

\subsubsection{Impact of Loss Coefficients}

\begin{figure}[]
    \centering
    \includegraphics[width=0.97\linewidth]{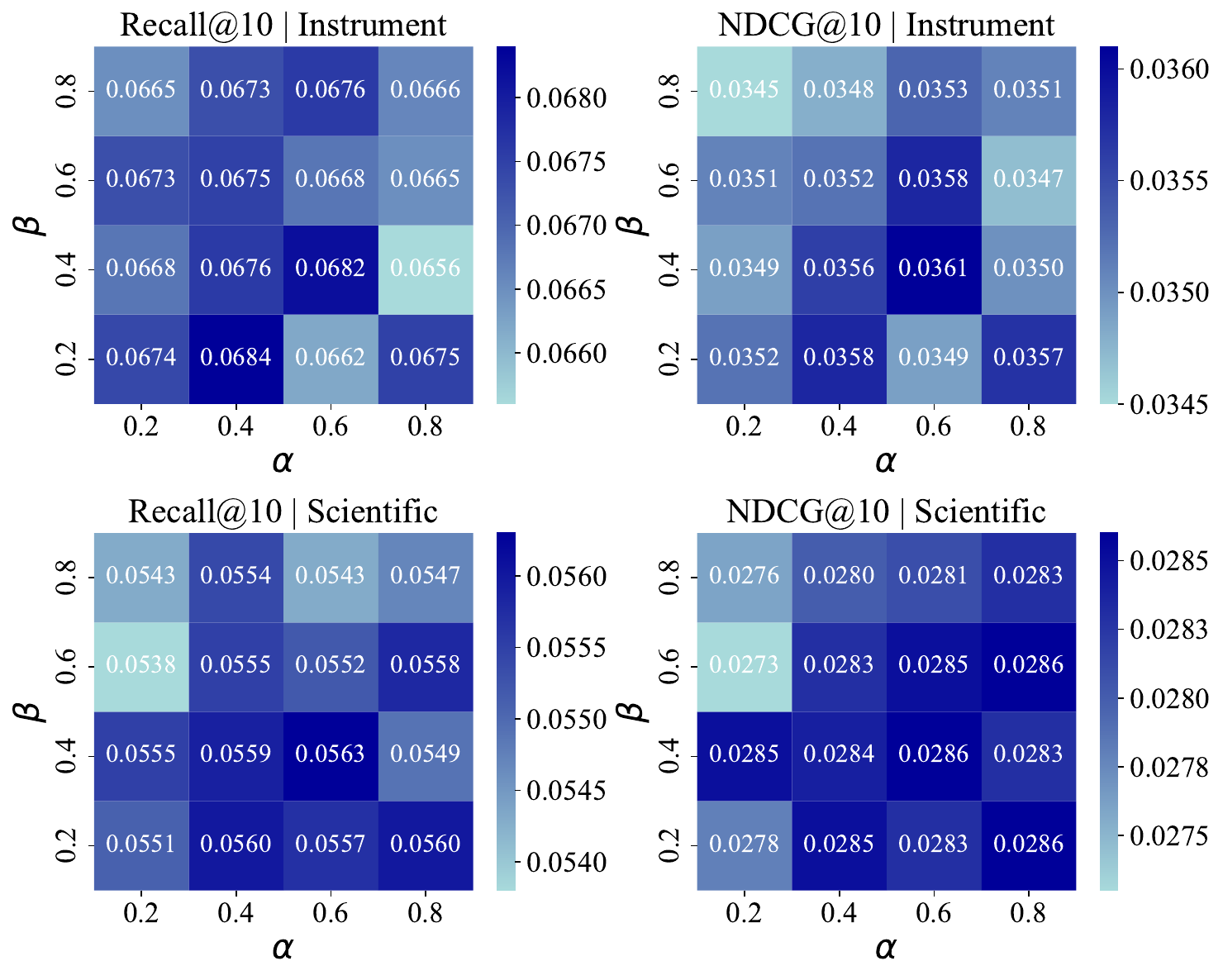}
    \captionsetup{font={small}}
    \caption{Performance comparison of different loss coefficients on Instrument and Scientific.}
    \label{fig:hyper_param}
\end{figure}

$\alpha$ and $\beta$ are key hyperparameters in CCFRec, controlling the importance of the masked code modeling (MCM) and masked sequence alignment (MSA) objectives. We evaluate their impact by varying their values within \{0.2, 0.4, 0.6, 0.8\} on the Instrument and Scientific datasets. The results for Recall@10 and NDCG@10 are shown in Figure~\ref{fig:hyper_param}.
On the Instrument dataset, CCFRec's performance initially improves but declines as $\alpha$ increases from 0.2 to 0.8, as moderate $\alpha$ values enhance semantic fusion while excessive values disrupt sequential pattern learning. On the Scientific dataset, the impact of $\alpha$ varies with $\beta$, indicating MCM's influence is less predictable. For $\beta$, its effect depends on $\alpha$, with the optimal value shifting based on $\alpha$, highlighting the need to balance both hyperparameters for optimal performance.

\subsubsection{Compatibility with ID-based Representations}
In previous sections, we excluded unique item IDs and relied solely on text embeddings and their associated semantic codes to model items and user preferences. However, our proposed model, CCFRec, is also compatible with ID-based representations. Below, we present a parameter-efficient approach for integrating unique item ID representations into our framework.
To begin with, we pre-train CCFRec using the original pipeline. Next, we introduce a raw item ID embedding table while freezing all other components of CCFRec. The unique ID embeddings are then summed up with the semantic-fused item representations and fed as input to the sequential recommender. The model is then fine-tuned according to the recommendation objective $\mathcal{L}_\text{CE}$, as defined in Eq.~\eqref{eq:ce}, until convergence. This fine-tuning process is highly efficient since only the item embeddings are trainable. The experimental results are summarized in Table~\ref{tab:add_id}.
The incorporation of unique item IDs further enhances the performance of CCFRec. This improvement can be attributed to the incremental collaborative information provided by unique IDs, which enable CCFRec to better distinguish between items. Unlike semantic codes, unique IDs offer a more flexible representation space, allowing for finer-grained differentiation.

\subsubsection{Analysis of the balance between collaborative and textual signals.}
\begin{figure}[]
    \centering
    \includegraphics[width=\linewidth]{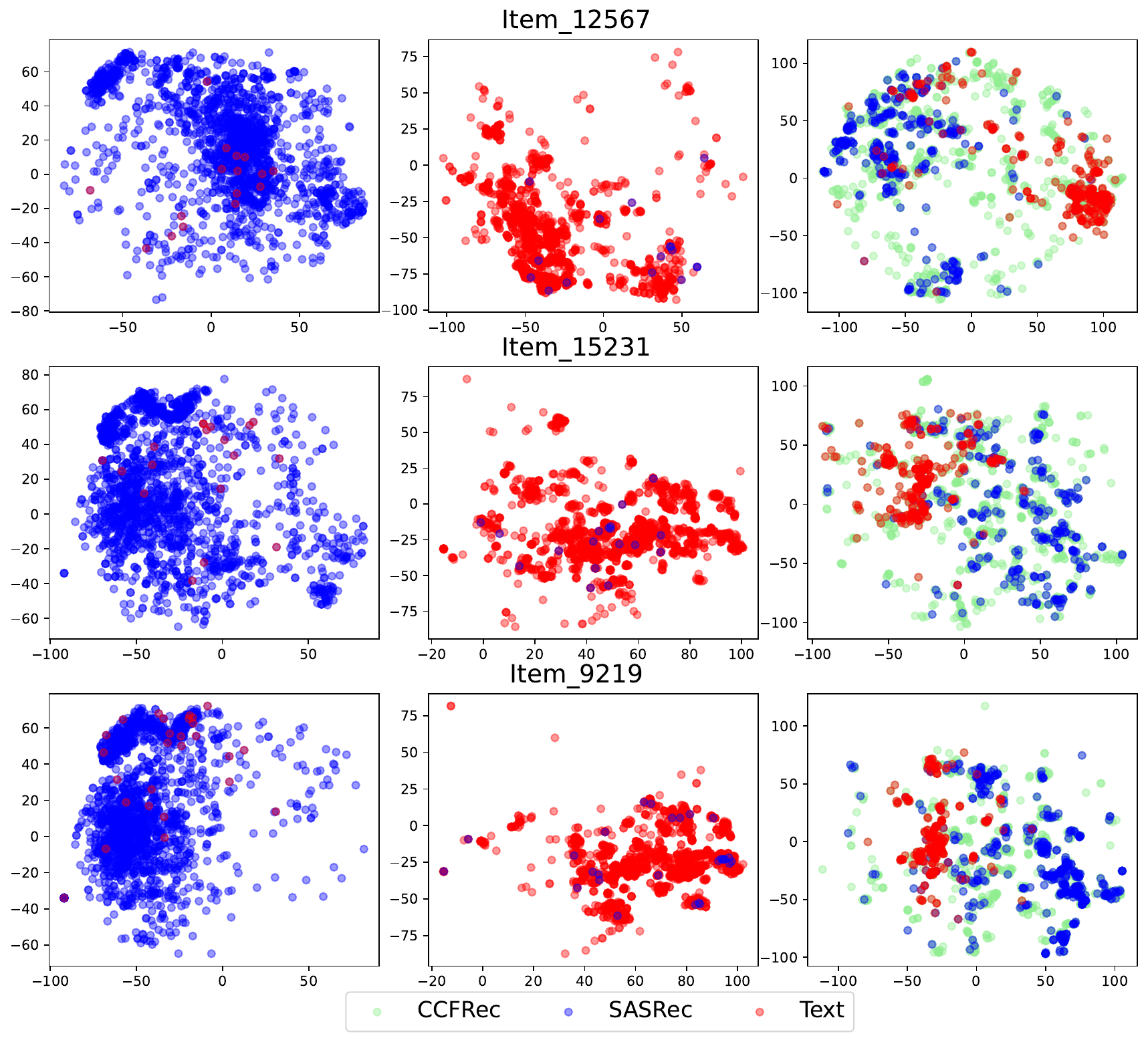}
    \captionsetup{font={small}}
    \caption{Visualization of similar items with different signals on Instrument. The left, middle, and right panels display t-SNE visualizations of the top-2000 similar items for a specific item derived from the SASRec embeddings, text embeddings, and CCFRec's fused representations, respectively.}
    \label{fig:tse}
\end{figure}

To validate our model's ability to bridge the textual and collaborative gap, we visualize its top-2000 similar items via t-SNE and use blue and red colors to represent the collaboratively similar and textually similar items, respectively. We present the examples of three items in Figure~\ref{fig:tse}.
The results reveal that CCFRec achieves a balanced distribution of textually and collaboratively similar items, confirming its ability to harmonize both information effectively. 

\begin{table}[]
\centering
\captionsetup{font={small}}
\caption{Performance of CCFRec empowered with ID-based representations on four datasets.}
\label{tab:add_id}
\resizebox{\columnwidth}{!}{%
\begin{tabular}{lrrrrrrrr}
\toprule
\multirow{2}{*}{Methods} & \multicolumn{2}{c}{Instrument} & \multicolumn{2}{c}{Scientific} & \multicolumn{2}{c}{Game} & \multicolumn{2}{c}{Baby} \\
\cmidrule(l){2-3} \cmidrule(l){4-5} \cmidrule(l){6-7} \cmidrule(l){8-9}
 & R@10 & \multicolumn{1}{c}{N@10} & R@10 & \multicolumn{1}{c}{N@10} & R@10 & \multicolumn{1}{c}{N@10} & R@10 & \multicolumn{1}{c}{N@10} \\ \midrule
CCFRec & 0.0682 & 0.0361 & 0.0555 & 0.0285 & 0.1042 & 0.0536 & 0.0455 & 0.0238 \\
+ Item ID & \textbf{0.0706} & \textbf{0.0366} & \textbf{0.0562} & \textbf{0.0288} & \textbf{0.1063} & \textbf{0.0540} & \textbf{0.0471} & \textbf{0.0246} \\
\bottomrule
\end{tabular}%
}
\end{table}

\section{Related Work}
\label{sec:related}

\subsection{Sequential Recommendation}
Sequential recommendation focuses on modeling user preferences from historical behavior sequences to predict the next item of interest. Early studies~\cite{fpmc, fossil} treated user behavior as Markov chains, emphasizing item transition relationships. Inspired by the success of deep learning in sequence modeling, researchers applied various neural network architectures for sequential recommendation, including CNNs~\cite{caser}, RNNs~\cite{gru4rec, DBLP:conf/recsys/TanXL16}, and GNNs~\cite{DBLP:conf/sigir/ChangGZHNSJ021, DBLP:conf/aaai/WuT0WXT19}. Recently, Transformer-based methods \cite{sasrec, bert4rec, DBLP:journals/tkde/HaoZZLSXLZ23} have shown significant success in sequential behavior modeling. Moreover, recent studies~\cite{rec-denoiser,convformer,ac-tsr,fame} have proposed to improve the Transformer architecture to further enhance sequential recommender performance. For example, FMLP-Rec~\cite{fmlp-rec} replaces the self-attention layer with filter-enhanced MLPs to reduce noise during user preference modeling. However, these methods rely solely on IDs to capture correlations among items, which leads to issues such as the cold-start problem.

\subsection{Text-Enhanced Recommendation}
With the rapid development of various application platforms, a growing amount of item content information has become available. This has motivated researchers to explore its potential for improving the performance of sequential recommender systems. We refer to this approach as text-enhanced recommendation. 
Previous studies~\cite{nova,dif-rec,kar,smlp4rec,tedrec,llmemb} have typically utilized text embeddings encoded by pre-trained language models (PLMs) to enhance item representations and improve behavior modeling. For instance, FDSA~\cite{fdsa} employs a dual-stream Transformer-based network to independently model item IDs and feature sequences. UniSRec~\cite{unisrec} uses an MoE-enhanced adaptor to extract universal information from text embeddings through multi-domain pretraining.
Another line of research~\cite{recjpq,kdsr,tiger,etegrec} utilizes the discrete codes derived from the text embeddings in sequential or generative recommendation. For instance, VQRec~\cite{vqrec} applies vector quantization to transform text embeddings into discrete item IDs, leveraging these quantized representations to model user preferences. MMSR~\cite{mmsr} uses clustering to decompose original modality features into discrete semantic codes, treating these indices as nodes to build a dense item relation graph.
Unlike the above methods, in CCFRec, we leverage discrete semantic codes as bridges to achieve effective textual-collaborative semantic fusion.

\section{Conclusion}
\label{sec:conclusion}

In this work, we proposed CCFRec to achieve effective textual and collaborative semantic fusion for sequential recommendation. The core idea of our approach is to bridge the textual-collaborative semantic gap through semantic codes. For this purpose, we construct multi-view text embeddings and corresponding semantic codes derived from vector quantization for fine-grained semantic utilization.
Next, we design a code-guided semantic-fusion module to enable flexible textual information extraction and integration. Furthermore, we devise an optimization method via code masking, comprising masked code modeling and masked sequence alignment, to model code correlations within individual items and enhance sequence modeling. Extensive experiments and in-depth analysis on four latest benchmarks demonstrate the effectiveness of our proposed CCFRec. In future work, we will leverage our method in other recommendation tasks (\eg multi-modal recommendation and multi-domain recommendation).
%%
% Acknowledgments.
\begin{acks}
This work was partially supported by National Natural Science Foundation of China under Grant No. 92470205 and 62222215, Beijing Municipal Science and Technology Project under Grant No. Z231100010323009, and Beijing Natural Science Foundation under Grant No. L233008. Xin Zhao is the corresponding author.
\end{acks}

\bibliographystyle{_acm/ACM-Reference-Format}
\balance
\bibliography{bibliography}

% \input{sections/6-appendix.tex}

%%
%% Appendix, this is the place to put it.
% \appendix

\end{document}